\def\ROSAT{{\it ROSAT\/\ }}
\def\ASCA{{\it ASCA\/\ }}
\def\SAX{{\it SAX\/\ }}

\def\ltsima{$\; \buildrel < \over \sim \;$}
\def\simlt{\lower.5ex\hbox{\ltsima}}
\def\gtsima{$\; \buildrel > \over \sim \;$}
\def\simgt{\lower.5ex\hbox{\gtsima}}

\def\@eprint{y}

\documentstyle[psfig]{mn}
\def\copyright{}

\title[A powerful, variable, off-nuclear X-ray source in NGC 4945]
{A powerful and highly variable off-nuclear X-ray source in the composite starburst/Seyfert 2 galaxy NGC 4945}

\author[W.N. Brandt, K. Iwasawa and C.S. Reynolds]
{\parbox[]{6.5in}{W.N. Brandt, K. Iwasawa and C.S. Reynolds}\\
\\
Institute of Astronomy, Madingley Road, Cambridge CB3 0HA\\
}

\begin{document}
\maketitle

\begin{abstract}  
We report the discovery of a powerful and variable off-nuclear X-ray source
in the nearby spiral galaxy NGC 4945. Two \ROSAT PSPC observations show the 
source to brighten in 0.5--2.0 keV flux by a factor of $\approx 9$ on a 
time-scale of 11 months or less. It is seen by \ASCA about one month
after the second PSPC pointing, and is seen to have dimmed by a factor of 
$\simgt 7$ in a \ROSAT HRI pointing about one year after the
second PSPC pointing. Its maximum observed 0.8--2.5 keV 
luminosity is $\simgt 8\times 10^{38}$ erg s$^{-1}$, making it brighter than
any known persistent X-ray binary in the Milky Way. Its total X-ray 
luminosity is probably larger than $1.2\times 10^{39}$ erg s$^{-1}$.
The observed variability argues against a superbubble 
interpretation, and the off-nuclear position 
argues against a low-luminosity active galactic nucleus. The source is 
therefore probably either an ultra-powerful X-ray binary or an ultra-powerful 
supernova remnant. Optical monitoring has not identified any supernovae
in NGC 4945 during the time of the X-ray observations, and any supernova
would have had to have been either very highly absorbed or intrinsically 
optically faint.
\end{abstract}


\section{Introduction} 

Recent X-ray studies of nearby spiral galaxies have revealed a wealth 
of activity (e.g. Fabbiano 1995 and references therein). Low-luminosity 
active galactic nuclei, starburst superbubbles of hot gas, and 
ultra-powerful X-ray binaries and supernova remnants are all currently 
subjects of widespread interest. Luminous and variable off-nuclear X-ray 
sources are of particular interest, as they often challenge our understanding
of accretion processes in stellar systems (e.g. Collura et~al. 1994; 
source A in IC 342, Okada et~al. 1994).
     
NGC 4945 is a nearby (3.7 Mpc: Mauersberger et~al. 1996), edge-on spiral 
galaxy in the southern hemisphere. It is the third brightest galaxy in 
the IRAS point source catalogue, and most of its infrared emission 
arises in a compact central region (Rice et~al. 1988; Brock et~al. 1988).
NGC 4945 shows both starburst emission (e.g. Heckman, Armus \& Miley 1990;
Koornneef 1993) and a highly obscured and 
variable Seyfert core (Iwasawa et~al. 1993). It
has recently been discovered to be the brightest Seyfert 2 known
at high X-ray energies ($\approx 100$ keV: Done, Madejski \& Smith 1996).
In this Letter we report on a powerful, off-nuclear X-ray
source in NGC 4945 that shows large-amplitude variability on a
time-scale of a year. 

\section{Observations and data analysis} 

\begin{figure*}
\centerline{\psfig{figure=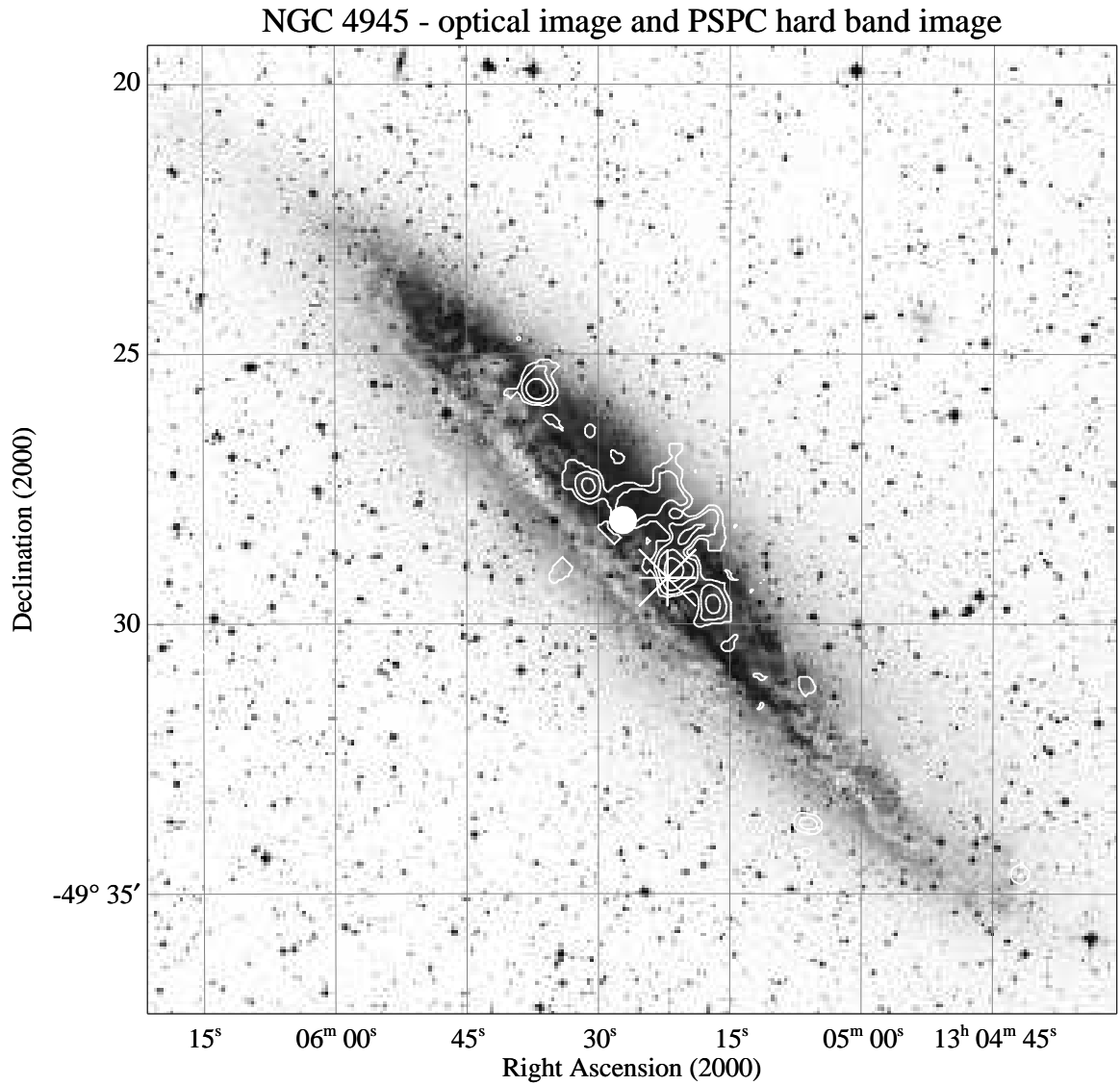,width=0.9\textwidth}}
\caption{Contours of the P2 \ROSAT PSPC X-ray image overlaid on the 
image from the UK Schmidt Southern Sky Survey J plate. Contours are at
8.8, 17.6, 35.2 and 70.6
per cent of the maximum pixel value. The location of the variable
X-ray source, NGC 4945 X-1, is marked with an eight-pointed star. The
nuclear position of Henkel et~al. (1994) is marked with a solid white dot. Note 
that NGC 4945 X-1 is significantly offset from the nucleus.}
\end{figure*}

NGC 4945 was observed with the \ROSAT 
Position Sensitive Proportional Counter (PSPC; Tr\" umper 1983; Pfeffermann et~al. 1987) starting on 
1992 August 12 (RP600259: total raw exposure of 14208 s over $2.66\times 10^5$ s) and 
1993 July 10 (RP600275: total raw exposure of 9032 s over 80760 s). These were the
first observations of this galaxy with an imaging X-ray detector. Hereafter we shall 
refer to RP600259 as `Pointing 1' or `P1' and RP600275 as `Pointing 2' or `P2'.
The \ROSAT observations were performed in the standard `wobble' mode,
and reduction and analysis of the PSPC data were performed with the
Starlink {\sc asterix} X-ray data processing system. NGC 4945 was also 
observed with \ASCA (Tanaka, Inoue \& Holt 1994) starting on 
1993 August 31 (total good SIS exposure time of 35000 s over 68362 s), 
and the \ROSAT High Resolution Imager (HRI) starting on
1994 July 11 (RH600276: total raw exposure of 8608 s over $6.27\times 10^5$ s).
The \ASCA data were reduced and analysed using the {\sc xselect} X-ray data 
processing system, and the HRI data were reduced and analysed with {\sc asterix}. 

In Figure 1 we show contours of the P2 PSPC hard image (we use PSPC channels 52--201
which correspond to $\approx$0.5--2.0 keV) 
overlaid on the image from the UK Schmidt Southern Sky Survey J plate  
[see section IIb of Lasker et~al. (1990) for more information on the
optical image]. We also show the position of the nucleus (we use the 
nuclear coordinates of Henkel, Whiteoak \& Mauersberger 1994).  
The \ROSAT image shows several point sources as well
as diffuse X-ray emission (presumably associated with the
starburst). We have used the {\sc asterix} point source 
searching program {\sc pss} (Allan 1995) to search the hard PSPC image for 
point sources that are spatially coincident with the optical extent of
NGC 4945. We detect five point sources (at greater than 5$\sigma$) in both P1
and P2. If we compare the count rates of these sources between P1 and P2,
four of them show little evidence for variability. However, the fifth
stands out as showing clear evidence for having 
brightened between P1 and P2 (see Figure 2).
We have plotted the position of this source in Figure 1 and shall hereafter
refer to it as NGC 4945 X-1. 
The P1 X-ray centroid position of X-1 is 
$\alpha_{2000}=$ 13$^{\rm h}$ 05$^{\rm m}$ 22.6$^{\rm s}$,
$\delta_{2000}=$ $-49^\circ$29$^{\prime}$12$^{\prime\prime}$, 
and its P2 X-ray centroid position is 
$\alpha_{2000}=$ 13$^{\rm h}$ 05$^{\rm m}$ 22.1$^{\rm s}$,
$\delta_{2000}=$ $-49^\circ$29$^{\prime}$09$^{\prime\prime}$. 
These positions each have absolute errors of about 15 arcsec and are
entirely consistent with one another. X-1 is offset from the
nucleus with extremely high statistical significance
(the separation is $\approx 1.3$ arcmin). We have checked the
PSPC image astrometry (for both P1 and P2) 
using NGC 4945A, which lies close to
NGC 4945 on the plane of the sky, and it appears to be good to
within 15 arcsec. The {\sc pss} count rate of X-1
during P1 is $(2.7\pm 0.6)\times 10^{-3}$ count s$^{-1}$, and
its {\sc pss} count rate during P2 is    
$(24.7\pm 2.3)\times 10^{-3}$ count s$^{-1}$ (we quote 68 per
cent confidence count rate errors). The hard \ROSAT band count rate 
thus appears to have increased by a factor of $9.1\pm 2.3$ in the 
11 months between P1 and P2. We note that, while small-scale gradients 
in the diffuse X-ray emission might lead to slightly larger errors
than the formal {\sc pss} errors stated above, they cannot remove the 
large-amplitude variability that we detect. Inspection by eye of the P1 and P2
images clearly shows that X-1 has brightened substantially. During P2, X-1 is 
the brightest soft X-ray point source observable in NGC 4945. 

\begin{figure}
{\psfig{figure=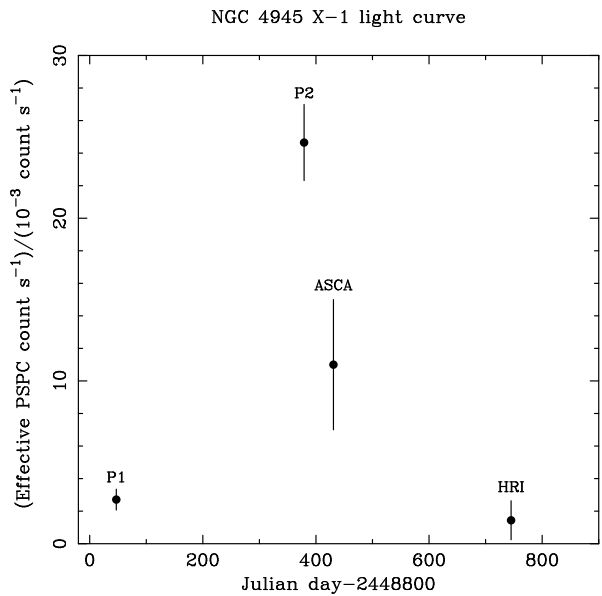,width=0.62\textwidth,angle=270}}
\caption{Effective PSPC count rate versus time for NGC 4945 X-1.
The effective PSPC count rates from \ASCA and \ROSAT HRI data
are computed as per the text.}
\end{figure}

The \ASCA observation also detects a point X-ray source with a position that 
coincides with that of X-1. We have used other sources in NGC 4945 seen by 
both \ASCA and \ROSAT to improve the astrometry of the \ASCA SIS images. 
The \ASCA X-ray centroid position for the source is  
$\alpha_{2000}=$ 13$^{\rm h}$ 05$^{\rm m}$ 22.9$^{\rm s}$,
$\delta_{2000}=$ $-49^\circ$29$^{\prime}$17$^{\prime\prime}$, 
and the error circle for this position is about 20 arcsec. The source is 
detected with very high statistical significance ($>6\sigma$).   
Precision source count rate determination is difficult, owing to the
broader \ASCA point spread function that leads to cross-source
contamination. In order roughly to estimate the count rate, 
we have compared the relative peak heights of X-1 and a
source located at 
$\alpha_{2000}=$ 13$^{\rm h}$ 05$^{\rm m}$ 31.9$^{\rm s}$,
$\delta_{2000}=$ $-49^\circ$27$^{\prime}$33$^{\prime\prime}$ 
in the 0.6--2.0 keV \ASCA image. The height of X-1 is larger than that of the 
comparison source by a factor of about 1.5. Provided that the comparison 
source is not variable we estimate an effective PSPC count rate for X-1 of 
about $11\times 10^{-3}$ count s$^{-1}$ by multiplying the PSPC count rate 
of the comparison source by a factor of 1.5. Thus X-1 appears to have
dimmed by a factor of about 2 in 52 days. We note that X-1 is
not the brightest point source in the 0.6--2.0 keV \ASCA image of 
NGC 4945.   
  
We have also used {\sc pss} to search the \ROSAT HRI image for point sources. In
this searching we use only HRI channels 4--16 to avoid contamination by 
ultraviolet light (cf. section 3.6 of David et~al. 1995). At the PSPC position
of X-1 there is only a weak (1.8$\sigma$) X-ray source with an HRI count
rate of $(0.45\pm 0.34)\times 10^{-3}$ count s$^{-1}$. The HRI count rate
of X-1 is thus a factor of $\simgt 28$ times lower than its P2 count rate. If we
use {\sc pimms} (Mukai 1995) and the PSPC spectral models given below
to take into account the different responses of the PSPC and the HRI, 
we find that the HRI count rate of X-1 should have 
only been lower than its P2 count rate by
a factor of 2.5--4.0 if X-1 remained constant. The HRI-to-P2 count
rate ratio of X-1 is much smaller than that 
for the other sources in NGC 4945. 
Thus X-1 appears to have dimmed by a factor of $\simgt 7$
between 1993 July 10 and 1994 July 11. 

Count rates from \ROSAT should be averaged 
over an integer multiple of the 400-s  
wobble period when used to search for rapid variability of cosmic 
X-ray sources (cf. Brinkmann et~al. 1994). We have searched for rapid 
variability of X-1 during P1 and P2 and do not detect any such variability
with high statistical significance. However, even 
during P2 the statistics are such that we would only be
able to detect variability by factors $\simgt 2$. The \ASCA
data do not show any significant X-1 variability (again variations
by a factor of $\simgt 2$ would be required for a detection),
and the HRI data do not allow a serious variability search. 

Meaningful PSPC spectral fitting of X-1 is not possible during P1, and only 
crude spectral fitting is possible during P2. Cross-source contamination, 
the low total number of source counts, and the limited spectral 
resolution of the PSPC all hinder precision fitting. However, in 
order to probe the spectrum of X-1 to the greatest extent possible and
derive luminosity constraints, we have extracted source counts from a circular 
region of radius 0.8 arcmin centred on X-1. In this region the emission from 
X-1 dominates the X-ray flux. While we recognize that we are losing
some source counts as a result of our small extraction radius, any larger radius
would introduce unacceptably large cross-source contamination. We do include 
the standard energy-dependent correction for 
source counts scattered outside our source region. We take our 
background from a circular region of radius 1 arcmin centred on  
$\alpha_{2000}=$ 13$^{\rm h}$ 05$^{\rm m}$ 10.8$^{\rm s}$,
$\delta_{2000}=$ $-49^\circ$31$^{\prime}$26$^{\prime\prime}$, 
as this region appears to have approximately the same level of background 
as the source region. We obtain $186\pm 15$ counts for our spectrum after 
background subtraction. The Galactic neutral hydrogen column towards NGC 4945 
is $(1.1\pm 0.3)\times 10^{21}$ cm$^{-2}$ (Heiles \& Cleary 1979, page 51),
and we note that there may well be significant intrinsic column in the disc 
of NGC 4945.  

After corrections for source region size, vignetting, wire scattering and 
detector dead time, we have binned our extracted spectrum so that there
are at least 12 counts in each spectral bin (this allows the use of
chi-squared fitting techniques). We fit this spectrum using the {\sc xspec}
spectral fitting package. We fit only the data above 0.85 keV, as the lower 
energy data have potentially significant background and cross-source 
contamination (by ignoring the data below 0.85 keV we lose only 
about 10 per cent of our source counts due to the large column).
When we state fluxes below they are computed for the 0.85--2.5 keV band
that we directly fit. All errors are for 90 per cent confidence,
taking all free parameters to be of interest other than the absolute
normalization. 
We first considered a power-law model with the 
only absorption being due to the Galactic column. 
This model gives a flat power-law photon index of 
$\Gamma=0.02\pm 0.7$,
$\chi^2_\nu=0.67$ for 12 degrees of freedom,
an absorbed flux of
$4.8\times 10^{-13}$ erg cm$^{-2}$ s$^{-1}$,
and an unabsorbed flux of 
$5.2\times 10^{-13}$ erg cm$^{-2}$ s$^{-1}$.
If we instead fix the photon index to be $\Gamma=2.0$ and allow the
absorption column to be a free parameter, we obtain
$N_{\rm H}=(7.0^{+2.9}_{-2.4})\times 10^{21}$ cm$^{-2}$, 
$\chi^2_\nu=0.74$ for 12 degrees of freedom,
an absorbed flux of
$4.0\times 10^{-13}$ erg cm$^{-2}$ s$^{-1}$,
and an unabsorbed flux of 
$8.0\times 10^{-13}$ erg cm$^{-2}$ s$^{-1}$.
Power-law models where we allow both the photon index and the absorption
column to be free parameters are very poorly constrained, but give 
unabsorbed fluxes similar to or larger than those stated above.
We have also fitted an absorbed thermal bremsstrahlung model to
our spectrum (see Figure 3). Using this model we derive
$N_{\rm H}=(6.0^{+8.0}_{-3.2})\times 10^{21}$ cm$^{-2}$, 
$kT=4.5$ keV ($kT$ is constrained to be larger than 0.5 keV),
$\chi^2_\nu=0.80$ for 11 degrees of freedom,
an absorbed flux of
$4.1\times 10^{-13}$ erg cm$^{-2}$ s$^{-1}$,
and an unabsorbed flux of 
$7.1\times 10^{-13}$ erg cm$^{-2}$ s$^{-1}$.
If we fix the total column at the Galactic value, a thermal bremsstrahlung
model leaves large systematic residuals and can be ruled out with
greater than 90 per cent confidence. 

The \ASCA cross-source contamination does not allow a detailed spectral 
analysis, but we have performed a rough X-ray colour analysis using a small 
aperture (a square with side length 1.2 arcmin) to minimize cross-source 
contamination. The colour analysis used three energy bands: 
0.7--1.3 keV, 1.3--2.3 keV and 2.3--8.0 keV. 
X-1 has the smallest 
2.3--8.0 keV to 1.3--2.3 keV count rate ratio ($0.84\pm 0.15$) 
as well as the smallest 
0.7--1.3 keV to 1.3--2.3 keV count rate ratio ($0.39\pm 0.07$)
of any source in NGC 4945. For a power-law model, the
first ratio corresponds to a spectral slope of $\Gamma=1.7$ or steeper, 
and the second ratio corresponds to an absorption column of
$5\times 10^{21}$ cm$^{-2}$ or more. This suggests 
that the Galactic column power-law fit to the
PSPC data (see above) is probably not physically appropriate. 
A more detailed colour analysis of all the sources in NGC 4945 will
be presented by Iwasawa et~al. (in preparation). 

\begin{figure}
{\psfig{figure=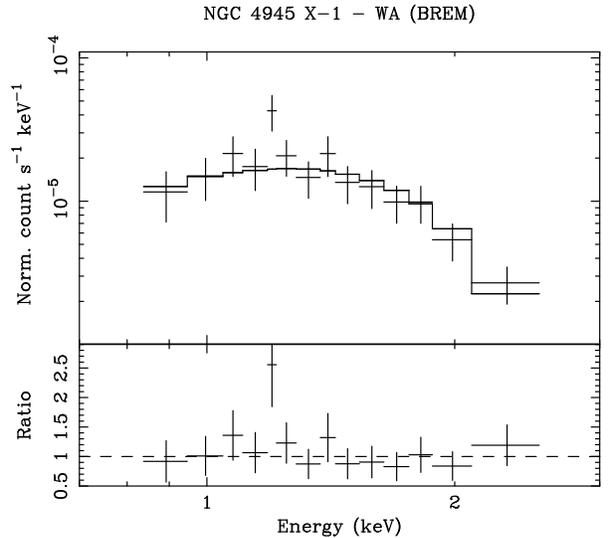,width=0.62\textwidth,angle=270}}
\caption{\ROSAT PSPC spectrum of NGC 4945 X-1 during P2. An
absorbed thermal bremsstrahlung model has been fitted to the 
data (see the text), and the data-to-model ratio is shown.}
\end{figure}
  
\section{Discussion} 

It is unlikely that X-1 is a foreground or background X-ray source.
Given the position of NGC 4945, the probability of finding a 
contaminating source with at least the P2 flux of X-1 within the optical 
extent of NGC 4945 is less than $1\times 10^{-2}$ (see section V
of Hertz \& Grindlay 1984). We have searched the UK Schmidt plate
image and the less saturated image in Sandage \& Bedke (1994), and
there are no obvious foreground optical counterparts. The position
of X-1 in the nomograph of Maccacaro et~al. (1988) does not agree
well with that of a foreground star (see their Figure 1). There is 
no strong X-ray variability during P2 as might be expected
from a foreground flare star and, in addition, \ASCA saw X-1 to be
bright 52 days after P2. Furthermore, we 
note that the excess column over the 
Galactic value suggested by our spectral analysis argues for 
the extragalactic nature of X-1. 

As all of our models above yield unabsorbed fluxes larger than 
$5\times 10^{-13}$ erg cm$^{-2}$ s$^{-1}$, the isotropic luminosity of
X-1 during P2 appears to be larger than 
$8\times 10^{38}$ erg s$^{-1}$ (we use a conservative distance of 
3.7 Mpc). We note that there is probably some additional flux outside 
the 0.85--2.5 keV band considered above, and reasonable estimates
of the total X-ray flux of X-1 suggest isotropic luminosities of  
$1.2\times 10^{39}$ erg s$^{-1}$ or more. 
There is no obvious radio source that corresponds to X-1
in the 843 MHz radio map of Harnett \& Reynolds (1985) or the 
CO maps of Dahlem et~al. (1993), although these limits are
not tight (e.g. the 843 MHz radio flux is less than 77 mJy with
a $43\times 57$ arcsec$^2$ beam). The large amplitude 
variability of X-1 allows us to rule out a starburst 
superbubble model for this source, and its off-nuclear position
argues against a low luminosity active galactic nucleus. We are thus
led to consider ultra-powerful X-ray binaries and supernova remnants.

The minimum P2 isotropic luminosity for X-1 corresponds to the Eddington 
limit luminosity for a 6 solar mass compact 
object, and makes it more luminous than any
known persistent X-ray binary in the Milky Way. If X-1 is a binary, then the 
variability and the mass derived from the Eddington limit are 
suggestive of a transient black hole binary, and we note 
that the peak observed isotropic luminosity of X-1 is similar
to that for the Galactic transient GS 2023+338 (V404 Cyg; Tanaka \& Lewin 1995). 
However, we note that some rare neutron star binaries in the Magellanic Clouds 
have comparable isotropic luminosities to X-1 (e.g. A$0535-668$ in outburst; 
SMC X-1 in its bright state). Of course, the emission need not be isotropic 
and indeed could be beamed (see Reynolds et~al. 1996 for a more detailed 
discussion of this possibility). Our X-ray spectrum is consistent to within
its (large) errors with that seen from luminous Galactic black hole
binaries.    

Ultra-powerful supernova remnants are described in Schlegel (1995)
and references therein. To explain X-1 a remnant would have had
to brighten by a factor of about 9 in one year and then dim by
a factor of $\simgt 7$ in the subsequent year. This could perhaps happen
if we caught a remnant either when it was very young or when it
happened to be interacting with circumstellar material. By comparison 
with predictions for the X-ray emission from the ring around 
SN 1987a (e.g. Suzuki, Shigeyama \& Nomoto 1992), we note that very 
dense circumstellar material would be needed in the second scenario 
to explain the large luminosity and short cooling time. However, 
the required variability does not seem to be obviously inconsistent 
with our limited knowledge of the variability of other powerful 
supernova remnants (Schlegel 1995). We have checked the 
updated version of the Asiago supernova catalogue and
the van den Bergh supernova catalogue
(Barbon, Cappellaro \& Turatto 1989; van den Bergh 1994), and 
no supernovae in NGC 4945 are listed. NGC 4945 
is regularly monitored for
supernovae, and no supernovae brighter than 15.5 mag have been
seen during the dates relevant to this 
paper (Rev. R. Evans, private communication; see chapter 5 of
Murdin 1990). We note that the `inclination effect' is
not likely to make the discovery of 
supernovae impossible in NGC 4945 (see the discussion in 
van den Bergh 1994). In the case of only Galactic extinction,
the apparent magnitude limit corresponds to an 
absolute magnitude limit of $-12.9$. If we
take into account the intrinsic extinction suggested by our X-ray
spectral fitting (see above), the absolute magnitude limit
is reduced to about $-15.6$, although limit values as low as $-20$ are
not impossible. However, we note that if the extinction were as
large as 7.1 magnitudes then even larger 0.85--2.5 keV luminosities
of $1.7\times 10^{39}$ erg s$^{-1}$ or more would be implied. Thus 
if X-1 were a supernova it probably, although not certainly, would 
have been seen at optical wavelengths. We have also checked 
the UK Schmidt plate database for optical 
plates that might be used to look for a supernova, but unfortunately 
there are no plates at appropriate dates.   

Additional {\it ASCA\/}, \ROSAT and \SAX observations of NGC 4945 are needed 
to monitor the variability of X-1. The detection of another outburst
or rapid X-ray variability would further support the binary
interpretation.  

\section*{Acknowledgments}

We gratefully acknowledge financial support from the United States 
National Science Foundation (WNB) and PPARC (KI, CSR).  
We thank H. Ebeling for the use of his {\sc imcont} image display
software, and R. Evans, A.C. Fabian, P. Meikle and M. Turatto for help. 
This research has made use of data obtained from 
the \ROSAT Data Archive Centers at the Department of Physics
and Astronomy, University of Leicester and Goddard Space Flight
Center. The UK Schmidt plate image was obtained via the 
Digitized Sky Survey which was produced at the 
Space Telescope Science Institute under United States Government 
grant NAG W-2166.

\bsp

\end{document}